\documentclass[]{jfm}

\usepackage{graphicx}
\usepackage{newtxtext}
\usepackage{newtxmath}
\usepackage{natbib}
\usepackage{hyperref}
\hypersetup{
    colorlinks = true,
    urlcolor   = blue,
    citecolor  = black,
}

\newcommand{\RomanNumeralCaps}[1]
\linenumbers

\title{DNS and role of round-off error: Two-Dimensional Taylor-Green vortex problem}

\author{Tapan K. Sengupta\aff{1}
  \corresp{\email{tksen@iitk.ac.in}},
 \and Vajjala K. Suman\aff{2,3}}

\affiliation{\aff{1} Professor (Retired), High Performance Computing Laboratory, IIT Kanpur, UP-208 016, India
\aff{2}Computational \& Theoretical Fluid Dynamics Division, CSIR-NAL, Bangalore, India
\aff{3}Department of Aerospace Engineering, IIT Kanpur, U.P.-208 016, India}

\begin{document}
\maketitle

\begin{abstract}
The role of round-off errors on the receptivity and instability of fluid flows are conclusively established for the first time using high accuracy simulations of the benchmark two-dimensional (2D) Taylor-Green vortex problem using double and quadruple precisions. Employing the fourth order Runge-Kutta (RK4) method for temporal discretization and Fourier pseudospectral method for spatial discretization enables unprecedented accuracy necessary for controlling all forms of errors, except the remaining round-off error. Results clearly show that adopting quadruple precision results in a qualitatively different receptivity route to turbulence and its subsequent decay compared to double precision. Another important observation is the identification of the receptivity phase which has never been reported before. Present study not only establishes the singular role of round-off errors but also has potential ramifications on receptivity and instability of flows due to precision of simulation.
\end{abstract}



\section{Introduction}
In scientific computing, errors are committed as one approximates various operators while discretizing and are called the truncation error. In solving the Navier-Stokes equation, one approximates the convection and diffusion operators by different methods. A good yard-stick has been adopted by expressing different methods in the spectral plane, as is customary (\citet{Vichnevetsky_Bowles}). It must be emphasized that the early practice of treating spatial and temporal discretizations separately are incomplete, and instead space and time discretizations are analyzed together for model convection equation (\citet{HACM}), convection-diffusion equation (\citet{Suman_CDE}) etc., and the generic approach is known as the global spectral analysis (GSA) which has been reviewed in \citet{GSA_2023}. With the help of GSA, it is now possible to understand the contributions of truncation error into explaining dispersion (phase errors, $q-waves$) and dissipation errors for various phenomenon including the Gibbs' phenomenon. GSA also  explains {\it anti-diffusion} which leads to catastrophic numerical instability shown by \citet{Suman_TKS_Mathur}. Both finite volume and finite element methods have been studied using GSA for convection equation in \citet{HACM}. In the context of representing spatial derivatives, the Fourier spectral method (\citet{Canuto_etal, Gottleib_Orszag}) is most accurate up to all resolved length scales i.e., Nyquist limit (\citet{HACM}).  

In contrast to the study of error dynamics due to truncation error, there are only a very few studies of error dynamics due to round-off error explained by GSA. The reason for this is due to the fact that in most situations truncation error dominates over round-off error. Thus, to study the effects of round-off error, one needs to carefully design a numerical study, as it is done here, to remove the other potential sources of errors, other than the round-off error. Another goal here is to establish whether the effects of the round-off error is epistemic or aleatoric. 

Round-off error is the difference of results obtained by an algorithm with exact arithmetic and that produced by the same algorithm using finite-precision. Thus, round-off error is due to truncated representation of real numbers during the arithmetic operations for numerical computations. It is conjectured that round-off error may accumulate, implying more refined calculations suffer more due to round-off error. It is noted that for ill-conditioned problems, such as the case of accumulating round-off error (\citet{Chapman, Chapra}). Thus in scientific computing, representation of real numbers with finite precision is one source of round-off error. Also, certain problems prone to physical/numerical instability can be influenced by such errors. For this reason, present investigation is performed on the disturbance growth for the Taylor-Green vortex (TGV) problem (\citet{Taylor_Green}). 

Existence of an analytical equilibrium solution prompted Taylor and Green to solve the TGV problem as a perturbation series in time to explain transition to turbulence. Goldstein also extended this perturbation series analysis by expanding it in terms of the Reynolds number ($Re$) - which is used as the inverse of the kinematic viscosity ($\nu$). The singularity in time and $Re$, prompted researchers to interpret singularities to imply turbulence.



 Due to the existence of an analytical periodical solution in 2D, one can use Fourier basis functions for spatial discretization (\citet{Orszag_1971, Orszag_1971a, Canuto_etal, Gottleib_Orszag, Brachet_etal1983, Brachet1991, Brachet_etal1992}) for enhanced accuracy. Fourier method is adopted to minimize truncation error in representing convection and diffusion terms, and initially \citet{MOF, Brachet_etal1983, Brachet_etal1992} numerically investigated the TGV problem by solving the Euler equation. \citet{Brachet_etal1983} also investigated the three-dimensional (3D) viscous flow problem for the generation of small-scale structures by vortex stretching in the resulting turbulence.
 


It is important to highlight the accuracy of the pseudo-spectral methods, in the context of the GSA to calibrate methods. Such methods are used in recent times, including the DNS of homogeneous isotropic turbulence by \citet{Buaria_etal2020}, who used $12288^3$ periodic grids. Such DNS uses a forcing term, following the work of \citet{Rogallo}, which also requires hyperviscosity for numerical stabilization (\citet{arxiv_DNS2021}). This is mandatory as it uses a two-stage Runge-Kutta (RK2) method for time advancement. The RK time integration method has been thoroughly investigated with the canonical convection and convection-diffusion equations by \citet{arxiv_DNS2021} which show that the RK4 method is significantly superior. Thus, there is no need to add numerical diffusion to eliminate numerical instability. Further details about these issues have been discussed in greater details for the double precision simulation of 2D TGV problem using Fourier spectral -RK4 method in \citet{EJMB_24}. The same methodology is used here for the double and quadruple precision simulations of the 2D TGV problem to investigate the role of precision on the receptivity.


The truncation error is absent, i.e. no dissipation and dispersion error, as the use of Fourier spectral method is equivalent to having a box filter. The use of 128 Fourier modes is adequate to also eliminate any error due to Gibbs' phenomenon. The only major source of error for the presented simulation is related to aliasing caused by the convection terms, and can be controlled by choosing correct de-aliasing techniques. For example, using the 3/2-rule of zero-padding of \citet{Canuto_etal}, one partially circumvents aliasing error. A correct zero-padding has been proposed with 2-rule in \citet{FCFD} that ensures complete removal of aliasing error.





The TGV problem in 2D or 3D form has allowed the study of the primary disturbance growth of the unsteady equilibrium flow by using near-spectral accuracy compact schemes by using nonuniform grids, reported in \citet{Sengupta_etal2DTGV, Gau_Hattori}. The 2D TGV problem has been investigated by \citet{Sengupta_etal2DTGV} with the help of the disturbance enstrophy transport equation (DETE). \citet{Sharma_Sengupta2019} report the vortex stretching term to accelerate the growth to the turbulent state in 3D by using a highly accurate compact scheme on a non-uniform grid. In contrast, the 2D form also displays the primary disturbance growth as shown in \citet{Sengupta_etal2DTGV}. While this primary growth for the 2D TGV problem is consistent with the perturbation analysis of \citet{Taylor_Green}, subsequently the Navier-Stokes solution displays the decay as has been theoretically shown for the monotonic decay of enstrophy by \citet{Doerring_Gibbon, ETE_2013}. This has been graphically demonstrated from the accurate numerical solution by \citet{EJMB_24} using Fourier spectral and RK4 method. This method is distinctly different from \citet{Brachet_etal1988} who used RK2 method along with hyperviscosity and persistent random excitation. The effect of hyperviscosity is so overpowering that the solution shows rapid relaminarization. Hence, to avoid this artifice, forcing has been often used on the right hand side of the Navier-Stokes equation in pseudo-spectral methods as shown in \citet{Rogallo, Eswaran_Pope}. Strictly speaking, such alteration of the governing equation does not qualify these methods to be termed as DNS. 

Here, we report for the first time the effects of computing precision on the dynamics of the flow and new insights into the physical instability via high accuracy simulations using RK4-Fourier pseudo-spectral method for the 2D TGV problem. This research is distinctly different from studies by other researchers because the present instability is triggered by implicit forcing by the round-off errors of the numerical scheme, instead of forced excitation, as is usually employed. As a result, receptivity is introduced into the problem and its effects are investigated thoroughly. 

The critical role played by precision in determining the success of engineering missions is documented in \citet{NASA_JPL,Patriot_missile_problem}. While the first reference documents precision as the singular factor leading to disaster with respect to missile defense systems, the second reference details the typical amount of digits of $\pi$ required by space mission planners for missions such as Voyager 1 to fulfill mission requirements. In CFD, double precision has been the de-facto norm for high performance computing and there have been some studies in the recent past where researchers concluded that precision does not induce observable effects as noted via simulations of the 3D TGV problem. 

A comprehensive view of the role of precision is established here by computing the 2D TGV problem for a $2\times2$ cell configuration, for a Reynolds number of $2000$, on a uniform grid with $128\times128$ points, using double and quadruple precisions. Apart from the difference in precision, all other parameters are identical including the time step. Thus, any observations or differences are directly attributed to precision used for computing. 

Two major highlights stem from the present work. First, a new phase of instability- ``the receptivity phase'' at the onset of computations, is identified and reported for the first time. Second, the effect of precision is shown to substantially delay the overall instability and alter disturbance structures even qualitatively. Further, round-off error is established as the singular reason in creating the seed for instability. As per our knowledge, such conclusions have not been established before.

\section{Governing equations \& Numerical Methods}
The 2D incompressible Navier-Stokes equations are solved here in streamfunction ($\psi$) - vorticity ($\omega$) formulation. Here, one solves an evolution equation for $\omega$ using the vorticity transport equation (VTE) and a Poisson equation for $\psi$ called the streamfunction equation (SFE). These equations are given below.

\begin{equation}
\frac{\partial \omega}{\partial t} + \frac{\partial (\omega u)}{\partial x} + \frac{\partial (\omega v)}{\partial y} = \frac{1}{Re}\nabla^2\omega
\end{equation}

\begin{equation}
\nabla^2\psi = -\omega
\end{equation}

The VTE is solved by evaluating the spatial derivatives using the Fourier basis and time integration using RK4 method. The SFE is solved using the Fourier basis itself which reduces to a very simple algebraic equation. Thus, iterative methods are not required for the SFE unlike finite difference or other methods.

\section{2D Taylor-Green Vortex Flow: Equilibrium and Linearized Disturbance Flow}
\label{sec:equl}
  


The TGV problem in a periodic domain $0 \leq (x,y) \leq 2\pi$ has the analytical solution

\begin{equation}
\begin{aligned}
\psi_m(x,y,t) &= \sin x\; \sin y\; F_m(t) \\ \omega_m(x,y,t) &= 2 \sin x\; \sin y\; F_m(t)
\end{aligned}
\label{eqAn}
\end{equation}
\noindent where $F_m(t)= e^{-\frac{2t}{Re}}$. This shows the analytic solution to decay exponentially with time. 

%
To analyze the effects of omnipresent disturbances, such as round-off errors, on the time-dependent equilibrium solution, we define $\psi = \psi_m + \epsilon \psi_d$ and $\omega= \omega_m + \epsilon \omega_d$, with subscripts $m$ and $d$ denoting the equilibrium and linearized disturbance components. From the requirement of spatial periodicity, $\psi_d$ and $\omega_d$ are given by,

\begin{equation}
\begin{aligned}
\psi_d(x,y,t) &= \sin x\; \sin y\; F_d(t) \\ \omega_d(x,y,t) &= 2 \sin x\; \sin y\; F_d(t)
\end{aligned}
\label{eqDis}
\end{equation}

Using the representations in Eqs. \eqref{eqAn} and \eqref{eqDis} in the SFE and the VTE, one notes the SFE to be an identity, and also the $O(1)$ equation of the VTE. The perturbation vorticity is obtained from the $O(\epsilon)$ VTE equation as 

\begin{equation}
    \frac{dF_d}{dt} = \frac{2F_d}{Re}
\end{equation}

This yields the disturbance vorticity as 

\begin{equation}
    \omega_d = \hat{F} e^{2t/Re} \sin x \sin y
\label{omd}
\end{equation}

\noindent with $\hat{F}$ as the disturbance vorticity at $t =0$. For very early times, one can assimilate this initial condition by expanding the exponential time dependent term by its simplified form at $t_\epsilon$ as $e^{2t_\epsilon/Re} \approx 1 + 2t_\epsilon/Re$. Thus, for such finite early times, $\omega_d$ will display a log-linear variation with time, which can be approximated as $\omega_{d,\epsilon} = t_\epsilon^{4\hat{F} \sin x \sin y / Re}$. A very accurate numerical simulation of the 2D TGV problem will be able to distinctly identify the two regimes of disturbance vorticity variation with time, as noted here. 
\section{2D TGV problem: Receptivity to decayed turbulence stage}
In the present research, the TGV problem is initialized using a $2 \times 2$-cell configuration in a domain given by $[0,2\pi] \times [0,2\pi]$. A uniform grid with $128 \times 128$ points is used and the Reynolds number $Re$ is chosen as $2000$. Double and quadruple precision simulations are run keeping all other numerical parameters the same. A time step of $0.025$ is chosen which gives very good accuracy according to analysis by  \citet{arxiv_DNS2021}. 

 


\begin{figure}
\begin{center}
\includegraphics[width=0.65\textwidth,keepaspectratio=true]{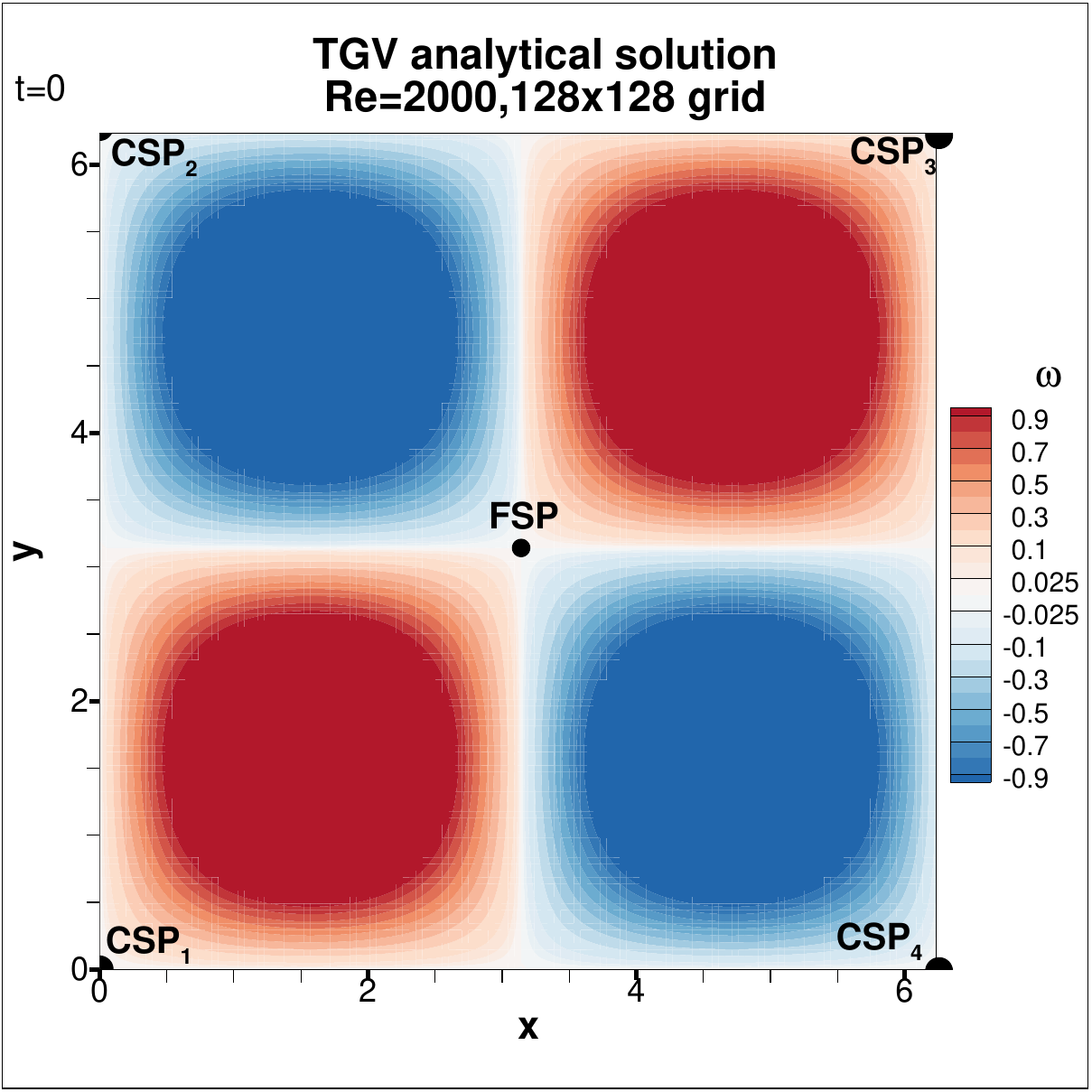}  
\end{center}
\caption{Free (FSP) and constrained (CSP) saddle points in the initial solution of the $2\times2$ cell 2D Taylor-Green vortex problem for a Reynolds number 2000 shown in the domain $0 \leq (x,y) \leq 2\pi$, for a $128\times128$ grid.}
\label{Fig1}
\end{figure}

In Fig. \ref{Fig1}, the initial analytical solution of the vorticity field is shown for the $2\times2$ cell pattern. This configuration forms saddle points which are precisely linked to the instability suffered by the problem. For the present configuration, we denote the saddle point in the center of the domain as free (FSP) whereas the corner ones are termed constrained (CSP) saddle points. The distinction is because the FSP is not influenced by periodic boundary conditions unlike the CSP and is the reason why varying the cell pattern display leads to different dynamics reported in \citet{EJMB_24}.   

\begin{figure}
\begin{center}
\includegraphics[width=0.65\textwidth,keepaspectratio=true]{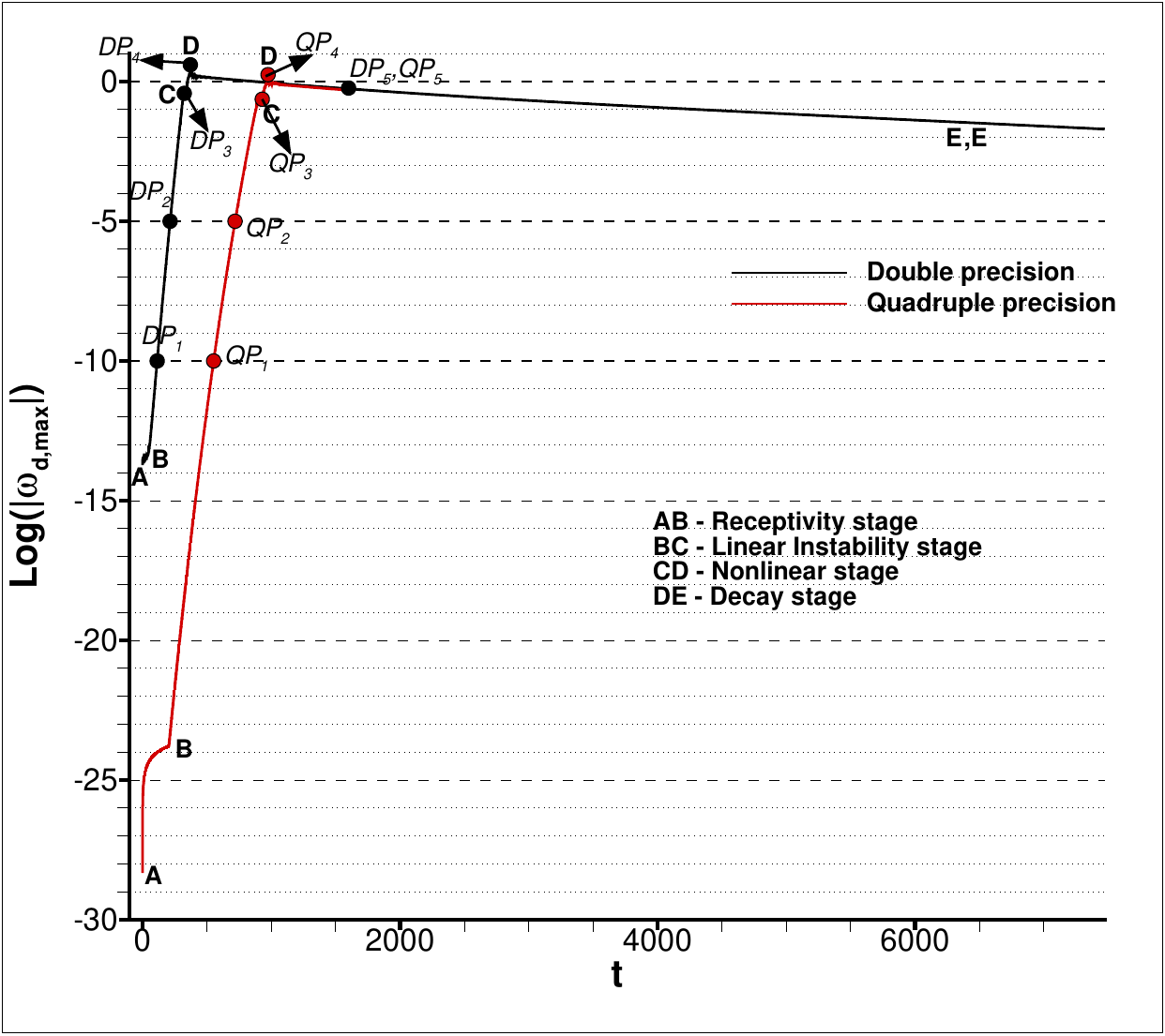}    
\end{center}
\caption {Phases of disturbance growth and the effect of precision for the 2D TGV problem shown via the evolution of the logarithm of the maximum of disturbance vorticity. Single character labels (A, B, C etc.) mark the onset/end of the growth phases. Labels DP and QP with subscripts denote the instants marked for analysis.}
\label{Fig2}
\end{figure}

In Fig. \ref{Fig2}, the time history of $Log(|\omega_{d,max}|)$ for the double (black) and quadruple (red) precisions is shown, where $\omega_{d,max}$ is the maximum disturbance in the domain showing a clear depiction of the differences due to precision. The analytical solution is initially perturbed by the round-off error. While the analytical solution decreases as $e^{-2t/Re}$, $\omega_d$ increases as given by Eq. \eqref{omd} which appears as the straight line BC in Fig. \ref{Fig2}. The most interesting aspect of the result is the portion AB which is known in the literature as receptivity (\citet{Sengupta21}). This refers to the conversion of background disturbances to log-linear growth indicated in Eq. \eqref{omd}. As this occurs at very early times, $\omega_{d,\varepsilon} = t_\varepsilon^{4\hat{F} \sin x \sin y / Re}$. This receptivity stage is more prominently noted for the quadruple precision compared to the double precision case.

Certain aspects of Fig. \ref{Fig2} are worth discussing with respect to the nature of the round-off error. If one were to consider the validity of the central limit theorem by \citet{Feller} caused by various sources that constitute the round-off error, then very large samples are required, which enable essentially random and independent data with finite variance. The probability distribution function, $p(x)$, is then Gaussian (normal). Then the probability $F(x)$ such that a random error lies between $-|x|$ and $+|x|$ is given by the error function,
$$F(x) = \frac{2}{\sqrt{\pi}} \int_0^{|x|/{\sqrt{2} \sigma}} e^{-t^2} dt = erf(\frac{|x|}{\sqrt{2} \sigma} )$$
\noindent where $erf$ denotes the error function; $\sigma$ is the standard deviation of the distribution and $1/[\sqrt{2\pi}\sigma]$ is the modulus of precision. 

It has been noted \citet{Ames} that when a number is rounded to $m^{th}$ decimal place, the error $\epsilon$ is never larger in magnitude than one half unit in the place of the $m^{th}$ digit of the rounded number, and the probability density function has a constant value, $ p = \frac{1}{2|\epsilon|_{max}}$ for $|\epsilon| < |\epsilon|_{max} = 5 \times 10^{-m-1}$; and is equal to zero elsewhere. This probability distribution can be poorly approximated by any normal distribution. By the central limit theorem, the cumulative distribution function of errors is approximately a linear combination of many such errors and is well approximated by a normal distribution. This is the motivation for treating round-off errors to be normally distributed.


From Eq. \eqref{omd}, it is apparent that the disturbance growth is monotonic, i.e. the growth is for zero frequency during the receptivity phase AB in Fig. \ref{Fig2}, for both double and quadruple precision. \citet{Sengupta_etal2DTGV} noted that the onset of the growth of disturbance occurs at the free saddle point, FSP. The cumulative action of the round-off error is due to forcing at multiple frequencies at the precision level, till the appearance of the zero frequency excitation from the interaction. Once this is latched upon via the action of receptivity, the exponential growth is log-linear, as noted in Fig. \ref{Fig2}. This unbounded linearized growth has been termed as singular (\citet{Taylor_Green}), and prompted many researchers to conjecture that it will lead to turbulence (\citet{Sengupta21}). The exponential growth rate is indicated by the slope given by the exponent $2/Re$. Numerical simulations (\citet{Sengupta_etal2DTGV, Gau_Hattori}) have not only identified the linear growth stage BC (in Fig. \ref{Fig2}), but also captured the subsequent non-linear growth CD. The resultant saturated 2D turbulent flow changes abruptly with an inflection point and starts oscillatory irregular damped fluctuations for both precision cases. Thereafter, one notes the monotonic decay of $Log(|\omega_{d,max}|)$, as theoretically shown in \citet{Doerring_Gibbon, ETE_2013} from the enstrophy transport equation. It is noted that \citet{EJMB_24} have shown a similar evolution of the multi-cellular 2D TGV problem (with $4 \times 4$ vortical cells) up to significantly longer time intervals (beyond $t = 5000$).

The growth stages shown in Fig. \ref{Fig2} indicate the distinct end of one phase and the beginning of a new phase for both double and quadruple precision. The receptivity stage (AB) indicates virtually a monotonic growth of $Log(|\omega_{d,max}|)$ at the respective precision levels. Phase BC displays a log-linear growth of $Log(|\omega_{d,max}|)$ with time. Beyond C, nonlinear growth sets in and continues up to D. It is interesting to note that beyond D, the effects of instability of the equilibrium flow disappear due to rapid distortion of the mean field. Thereafter, the effects of the enstrophy transport equation by \citet{Doerring_Gibbon, ETE_2013} take over, depicting the linear decay for both double and quadruple precision simulations.

In Fig. \ref{Fig2}, we have plotted $Log(|\omega_{d,max}|)$ without mentioning the location where this occurs. It was noted earlier that the onset of disturbance growth is at the free saddle point, after which the effects percolate to other points in the domain (\citet{Sengupta_etal2DTGV}). To understand the cumulative differences of the disturbance growths for double and quadruple precision simulations, time instants are chosen and marked with labels $DP_i$, $QP_i$, $i$ being a subscript, in order to compare the flow fields during the identified phases. 

\begin{figure}
\centering
\includegraphics[width=1\textwidth]{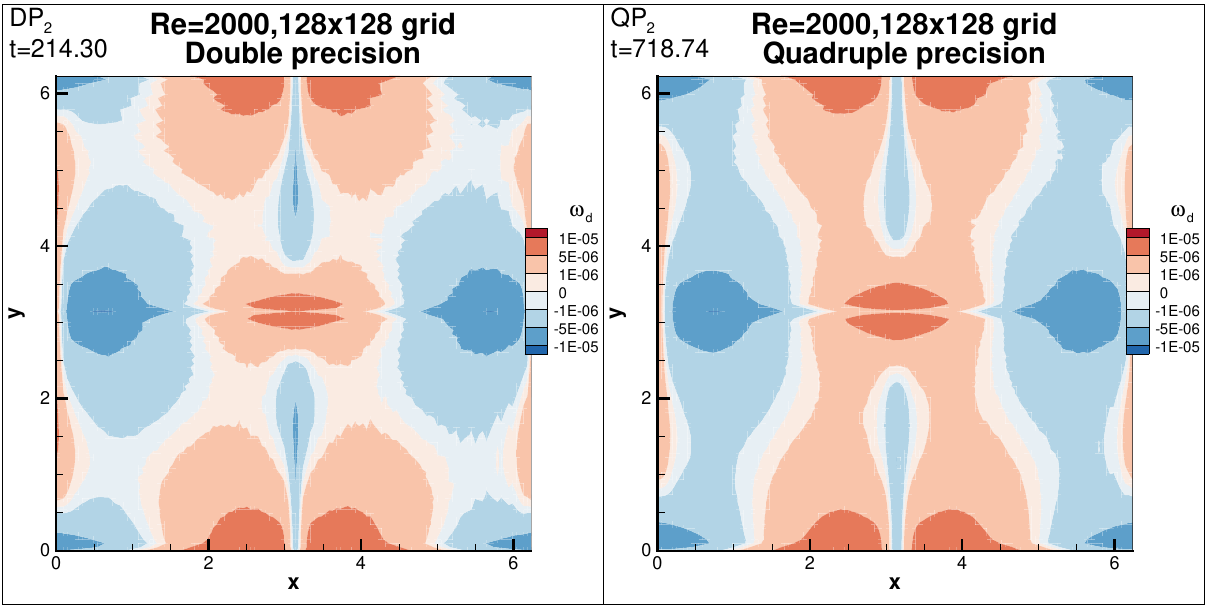}
\caption  {Comparison of the disturbance vorticity fields between double and quadruple precision for the same tolerance level of $10^{-5}$ ($DP_2$ and $QP_2$) obtained for $Re=2000$ and $128\times128$ grid.}
\label{Fig6}
\end{figure}

In Fig. \ref{Fig6}, the $\omega_d$ fields are compared between double and quadruple precisions for the instant when $Log(|\omega_{d,max}|)=10^{-5}$ i.e., $DP_2$ and $QP_2$, respectively. No significant differences are observed despite a large time difference and different excitation of round-off errors. The vortical structures show qualitatively similar levels, except the negative regions which are more developed for the quadruple precision case. 

\begin{figure}
\centering
\includegraphics[width=0.85\textwidth]{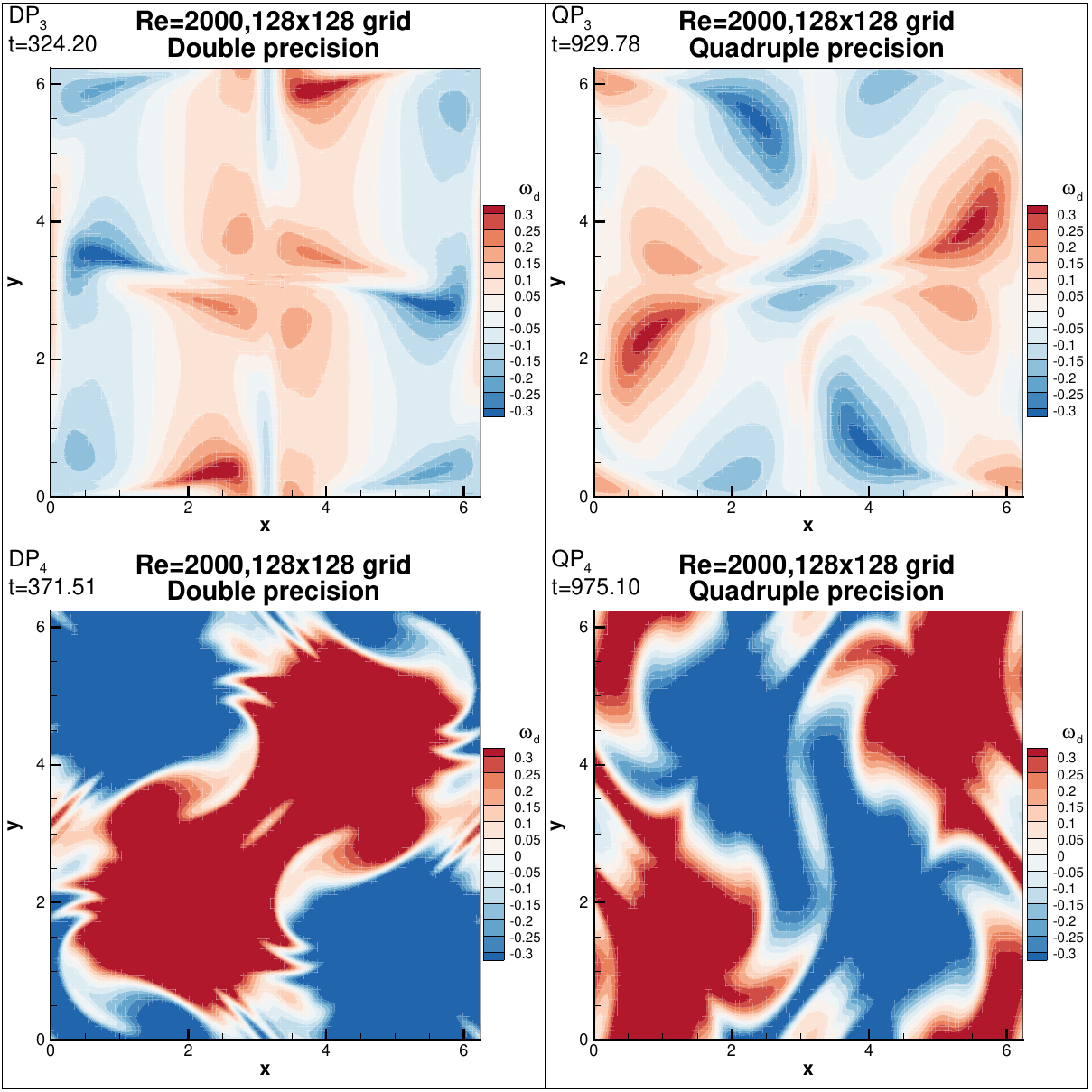}
\caption  {Comparison of the disturbance vorticity fields between double and quadruple precision at the commencement of the nonlinear phase (top) and maximum instability state (bottom) obtained for $Re=2000$ and $128\times128$ grid.}
\label{Fig7}
\end{figure}

In Fig. \ref{Fig7}, $\omega_d$ fields are compared between double and quadruple precision computations for the inception of the nonlinear stage ($DP_3$ and $QP_3$) and the maximum state of instability ($DP_4$ and $QP_4$), respectively. Retention of some structures at earlier times during the linear exponential growth stage implies that the flow is still related to the analytical solution. This is distinctly noted in Fig. \ref{Fig6} and the top frames of Fig. \ref{Fig7}. However, the bottom frames for the peak disturbance growth at D show no such correlation with the analytical solution.

There is another important aspect to note from the dynamics of the TGV problem. For the time duration depicted here, the major players dictating the flow evolution are the receptivity of the flow to round-off error; the linear / nonlinear stage of disturbance growth followed by the subsequent decay stage, where the enstrophy transport plays an important role. Apart from the receptivity stage, there is a competition between the effects of the disturbance growth via nonmodal global instability (established for wall bounded flows by \citet{nonmodal_PRR}) and the evolution of enstrophy given by its transport equation in \citet{Doerring_Gibbon, ETE_2013}. Researchers have tried to study the juxtaposition of these two physical mechanisms via the disturbance enstrophy transport equation by \citet{Asengupta_etal2018}. 

An interesting aspect to highlight from the present research is that the nonmodal growth is monotonic up to the maximum energy/ enstrophy stage at D. Thereafter the growth saturates and the sole mechanism of enstrophy transport equation overrides it. This is similar to the observations for the flow over a flat plate at zero incidence, where the nonmodal growth causing transition to turbulence by the creation and growth of spatio-temporal wave front (STWF) saturates after fully developed turbulence. A similar maximum has been noted earlier by other researchers for 3D TGV (\citet{Brachet_etal1983}) and zero pressure gradient boundary layer (\citet{Sengupta21}). There is a decaying turbulence stage between D and E, shown in Fig. \ref{Fig2}. Beyond E, there are no perceptible fluctuations and the flow can be viewed strictly as laminar. This is according to the enstrophy transport equation for 2D periodic flow by \citet{Doerring_Gibbon, ETE_2013} which shows the enstrophy $\Omega=\omega\cdot\omega$ to decay monotonically as $\frac{D\Omega}{Dt} = -\frac{2}{Re} (\nabla \omega)^2$. 

\section{Summary and Conclusions}
\begin{itemize}
    \item The role of round-off errors and precision on receptivity is conclusively established for the first time by careful 2D simulations of the Taylor-Green vortex problem with double and quadruple precision. 
    \item Simulations are performed with RK4-Fourier pseudo-spectral method which provide unprecedented accuracy in order to analyze effects of round-off errors on the receptivity.
    \item Round-off errors play a singular role in the ensuing spatio-temporal vorticity dynamics as the higher quadruple precision resulted in a significant delay in the creation and evolution of receptivity compared to the double precision case.
    \item A receptivity phase (AB) is identified for the first time for both precision cases, denoting the internalization of the round-off errors in the flow. This creates an environment which influences the subsequent stages of flow evolution as opposed to the commonly accepted perception that it is activated right from the initial state.
    \item Differences in precision are evident during the nonlinear stage CD and their investigations will be reported later. 
\end{itemize}



\backsection[Funding]{This research received no specific grant from any funding agency, commercial or not-for-profit sectors.}

\backsection[Declaration of interests]{The authors report no conflict of interest.}

\bibliographystyle{jfm}
\bibliography{ROE}

\end{document}